\documentclass{JHEP3}
\usepackage{amsmath}
\usepackage[pdftex]{graphicx}
\usepackage{epsfig,multicol,bbm}

\newcommand{\lumi}{\;\mbox{cm}^{-2}\mbox{s}^{-1}}
\newcommand{\e}{\mathrm{e}}
\title{Diffractive $\Upsilon$ production at the Tevatron and LHC}
\author{B.E. Cox, J.R. Forshaw \\
School of Physics \& Astronomy, University of Manchester, \\
Oxford Road, Manchester M13 9PL, U.K.\\
\email{brian.cox@cern.ch}, \email{jeff.forshaw@manchester.ac.uk} }
\author{R.Sandapen \\
D\'epartement de Physique et d'Astronomie, Universit\'e de Moncton, \\
Moncton, N-B. E1A 3E9, Canada.\\
\email{ruben.sandapen@umoncton.ca} }
\preprint{MAN/HEP/2008/45}
\abstract{
We compute the rate for diffractive $\Upsilon$ meson production at the Tevatron and the LHC.
The $\Upsilon$ is produced diffractively via the subprocess
$\gamma+p \to \Upsilon+p$ where the initial photon is radiated
off an incoming proton (or antiproton). We consider the possibility to
use low angle proton detectors to make a measurement of the
$\gamma p$ cross-section and conclude that a measurement of
the cross-section at a centre of mass energy in excess of 1~TeV is
possible at the LHC. This is in the region where saturation effects are likely to
reveal themselves. 
}

\begin{document}
\section{Introduction}
\begin{figure}[h]
\begin{center}
\includegraphics[width=2.7in]{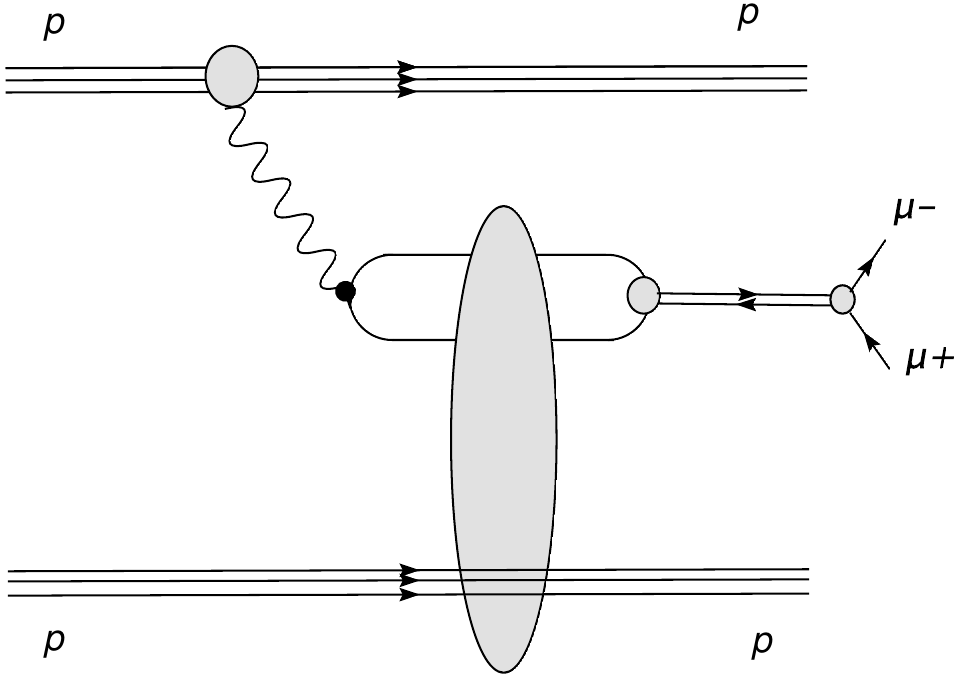}
\end{center}
\caption{Diffractive photoproduction of the $\Upsilon$ meson in $pp$
  collisions. At the Tevatron one $p$ stands for an antiproton and the other for a proton while at the LHC, both stand for protons. The $\Upsilon$ decays into a $\mu^+\mu^-$ pair which is
  detected. At the LHC, there is opportunity to also detect one of the protons.}
\label{fig:process}
\end{figure}

In this paper we calculate the rate for the reaction $p+p\rightarrow
p+ \Upsilon(nS) + p$ where $n=1,2,3$. The $\Upsilon$ is produced diffractively
after one of the incoming protons radiates a photon, as
illustrated in Figure~\ref{fig:process}, i.e. via the sub-process
$\gamma+p\rightarrow \Upsilon+p$.

The reaction is interesting primarily because the LHC will probe
values of the $\gamma p$ centre-of-mass energy well beyond those
that were reached at HERA. As a result, there is an opportunity to examine QCD in
the region where non-linear (saturation) effects are expected to be
important. In addition, the odderon is expected to contribute at some
level and an accurate measurement would help to constrain the theory. We shall not consider the odderon in this paper. 

Ordinarily this is a measurement that could only be
performed at low LHC luminosities ($\sim 10^{31}\lumi$) in order to
avoid problems with pileup. However, there is interest
in supplementing the LHC general purpose detectors with low angle
proton detectors at 420~m from the interaction points; the primary goal being the study of `central
exclusive' production of new physics, and in particular the reaction
$p+p\rightarrow p+H+p$ \cite{Higgs1}. 
The $\Upsilon$ states are too light to be
produced in conjunction with two measured protons but measurement of
one proton is feasible and that may be 
sufficient to control pileup induced backgrounds at
instantaneous luminosities in excess of ($10^{33}\lumi$). Work is currently in progress to measure
exclusive $\Upsilon$ production at the Tevatron \cite{CDF:DIS08} and
the CDF collaboration has identified centrally produced
charm meson states \cite{CDF:charm} and centrally produced dileptons
\cite{CDF:dileptons}. 

There already exist a number of papers
predicting the rate for this process \cite{KN:2004,MW:2008,RSS:2008,Odderon,Ivanov:2004nk,Goncalves:2007sa}. We add to what has been done by
providing predictions using the dipole model cross-section of
\cite{FS:2004} that successfully describes a wide range of the HERA
data \cite{Forshaw:2006np}. We also make an estimate of the effect of
detector acceptances and the impact of measuring one of the protons.   

In the following section we explain how to calculate the rate for this process. The
flux of photons off an incoming proton is well known, as is the dipole
cross-section that determines how the photon produces the $\Upsilon$  after first
fluctuating into a $\bar{q}q$ pair. The latter is well
determined\footnote{At least for light quark scattering.}, primarily as a result
of high quality data on the deep inelastic structure function $F_{2}$
measured at HERA \cite{H1F2:01,ZEUSF2:01}. In Section 3, we compute the photoproduction cross-section
and compare our predictions with the available HERA data. In Section
4, we present our
predictions for the rapidity distributions of the $\Upsilon$ and
the expected rates at the Tevatron and the LHC, including the effects
of proton tagging.

\section{The $pp$ cross-section}
Entirely in analogy to photoproduction at HERA, where almost on-shell photons
are radiated off incoming electrons according to the Weisz\"{a}cker-Williams
distribution, protons at the LHC can radiate almost real photons that can
then scatter off protons heading in the opposite direction. A significant
fraction of these $\gamma p$ reactions will be diffractive and leave the
incoming proton intact. Amongst these will be the process
$\gamma+p\rightarrow V+p$ where $V$ is a vector meson and that is
where our interest lies. 

If the $\Upsilon$ is produced at rapidity $Y$ then its
energy $E_{\Upsilon}\approx M_\Upsilon \cosh Y$ (we neglect the
transverse momentum of the meson). Defining 
\begin{equation}\xi = \frac{M_\Upsilon}{\surd s} \e^Y 
\end{equation}
where $\sqrt{s}$ is the center-of-mass energy of the $pp$ system,
the cross-section of interest is
\begin{equation}
 \frac{\mathrm{d}^2 \sigma(pp\rightarrow p \Upsilon p)}{\mathrm{d} Y
   \mathrm{d} Q^2} = \xi f_{\gamma
/p}(\xi,Q^2)\;\sigma_{\gamma p}(\xi s) + (~Y \to -Y~)~,
\label{pp-ddifxsec}
\end{equation}
where $f_{\gamma/p}(\xi,Q^2)$ is the photon flux given by \cite{Drees:1988pp}:
\begin{equation}
f_{\gamma/p}(\xi,Q^2)=\frac{\alpha_{{\mathrm{em}}}}{2\pi}\frac{1+(1-\xi)^{2}}{\xi}\frac{1}{Q^{2}}
\frac{1}{\left(  1+Q^{2}/\mu^{2}\right)  ^{4}}~,
\label{photon-flux}
\end{equation}
and $\mu^2=0.71$ GeV$^{2}$ fixes the electromagnetic form factor of the
proton. $\xi$ is the fractional energy loss of the proton. The cross-section for $\gamma+p \to \Upsilon + p$ is denoted by
$\sigma_{\gamma p}(W^2)$ where $W$ is the $\gamma p$ centre of mass
energy, i.e $W^2 = M_{\Upsilon} \sqrt{s} \exp({\pm Y})$. $W$ is not uniquely defined by the $\Upsilon$'s rapidity since any one of the two incoming protons can radiate the photon. 
The two terms on the right-hand-side of Eq.~(\ref{pp-ddifxsec}) allow for both possibilities.\footnote{By adding cross-sections, we are neglecting the interference between the two amplitudes.}

The minimum virtuality of the photon is
\begin{equation}
Q_{\text{min}}^{2}  \approx \xi^2 m_{p}^2~,
\label{Q2min}
\end{equation}
where $m_{p}$ is the proton mass.
Integrating Eq.~(\ref{photon-flux}) over $Q^2$ yields the integrated flux \cite{Drees:1988pp}:
\begin{equation}
f_{\gamma/p}(\xi)=\frac{\alpha_{\mathrm{em}}}{2\pi}\frac{1+(1-\xi)^{2}}{\xi}\left(  \ln
A(\xi)-\frac{11}{6}+\frac{3}{A(\xi)}-\frac{3}{2A(\xi)^{2}}+\frac{1}{3A(\xi)^{3}}\right)
\end{equation}
with
\begin{equation}
A(\xi)=1+\mu^2/Q_{\text{min}}^{2}
\end{equation}
and thus the rapidity distribution of the $\Upsilon$ is given by
\begin{equation}
\frac{\text{d}\sigma(pp\rightarrow p\Upsilon p)}{\text{d}Y}=\xi f_{\gamma
/p}(\xi)\text{ }\sigma_{\gamma p}(\xi s) + (~ Y \rightarrow -Y~) ~.
\label{pp-difxsec}
\end{equation}

So much for the photon flux, we are now ready to tackle the cross-section for
$\gamma+p\rightarrow \Upsilon + p$.  If the invariant mass $W^{2}\gg M_{\Upsilon}^{2}$ then we
are in the diffractive regime and consequently the imaginary part of
the forward $(t=0)$ amplitude can be approximated by\footnote{We assume that
  the $\Upsilon$ retains the helicity of the photon.}:
\begin{equation}
 \Im \text{m}\; \mathcal{A}_{\gamma p}(W^2)=\frac{1}{2}{\displaystyle\sum\limits_{\lambda,h,\bar{h}}}
\int\text{d}^{2}r\text{ d}z\text{ }\psi_{\lambda,h\bar{h}}^{\Upsilon \ast}(z,r)\psi_{\lambda,h\bar{h}}^{\gamma}(z,r) \sigma(x,r)~,
\label{factorized-amplitude}
\end{equation}
where $\psi_{\lambda,h\bar{h}}^{\gamma}(z,r)$ and $\psi_{\lambda,h\bar{h}}%
^{\Upsilon}(z,r)$ are the light-cone wavefunctions of the photon and the $\Upsilon$ (of
helicity $\lambda$). They represent the amplitude for the vector meson to
fluctuate into a $b\bar{b}$ pair of transverse size $r$ and with the quark
carrying energy fraction $z$ and helicity $h=\pm\frac{1}{2}$. We take $x=M_{\Upsilon}^2/W^2$. 

The cross-section $\sigma(x,r)$ is called the dipole
cross-section and it represents the cross-section for the $b\bar{b}$ pair to
scatter off a proton. The formalism we are describing has been very
successful in explaining a very wide range of HERA data, including the
extremely precise data on the $F_{2}$ structure function at low $x$
\cite{Forshaw:2006np}. The dipole cross-section has been extracted,
for light-quark dipoles, from HERA data and we use a parameterization
introduced in \cite{FS:2004}; we shall turn to this shortly. First we
address the calculation of the light-cone wavefunctions. 

The photon wavefunction is well known and, since the longitudinal
wavefunction of a real photon vanishes, we shall only need the
transverse ($\lambda=\pm 1$) photon wavefunction given by 
\begin{align}
\psi_{\lambda,h\bar{h}}^{\gamma}(z,r)= &  \sqrt{\frac{N_{\text{C}}}{4\pi}%
}\sqrt{2}\frac{ee_{f}}{2\pi}\{\delta_{h,\bar{h}}\delta_{\lambda,2h}m_{b}%
K_{0}(m_b r)\nonumber\\
& -i(2h)\delta_{h,-\bar{h}}\text{e}^{i\lambda\phi}\left[
(1-z)\delta_{\lambda,-2h}+z\delta_{\lambda,2h}\right] m_b K_{1}(m_b r)\}~,
\end{align}
where $e^2 = 4\pi \alpha_{\mathrm{em}}$ and $e_{f} = -1/3$ is the electric
charge of the $b$ quark.

\begin{figure}
 \centering
\includegraphics[width=4cm]{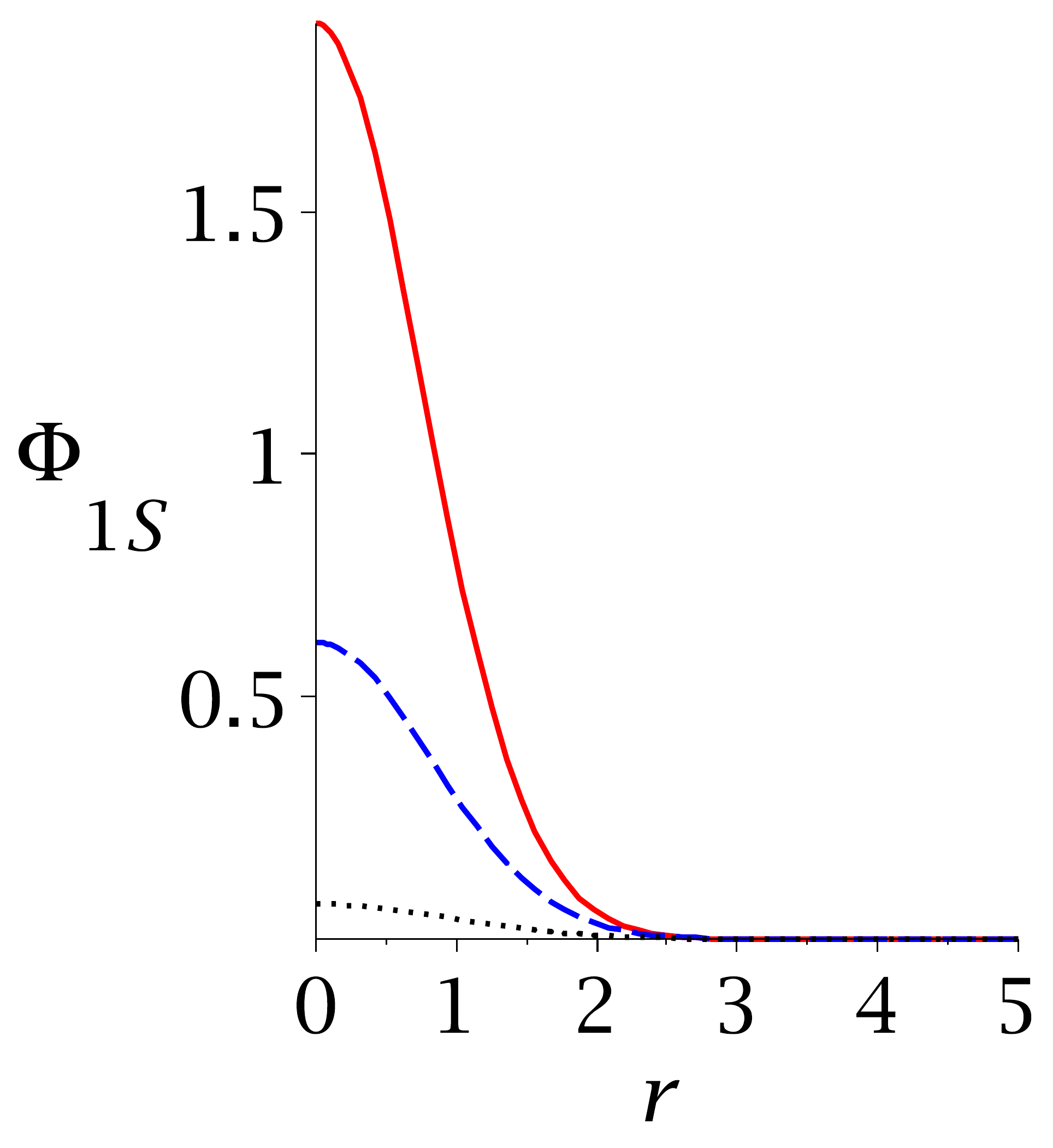}\hspace{0.1cm}\includegraphics[width=4cm]{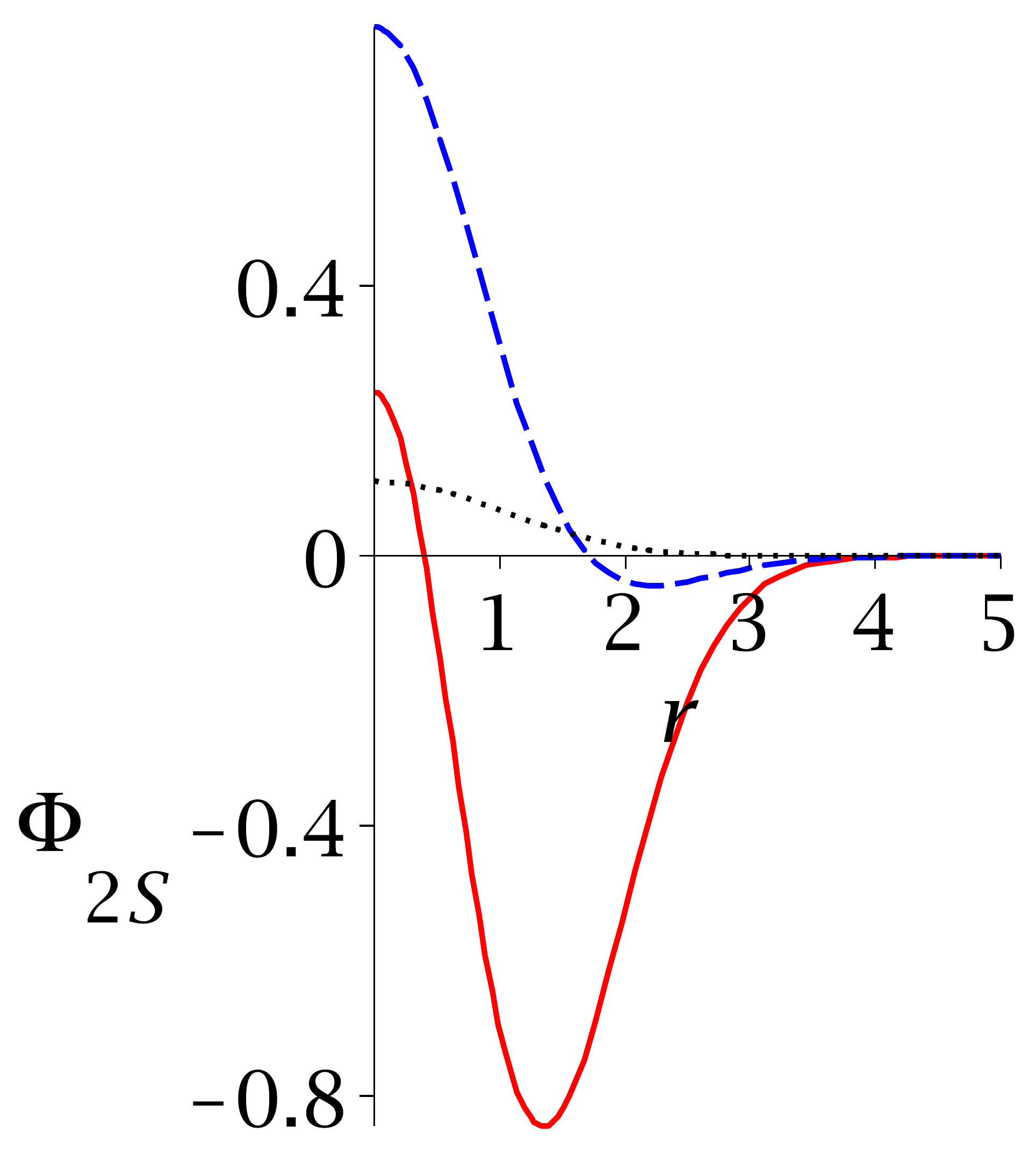}\hspace{0.1cm}\includegraphics[width=4cm]{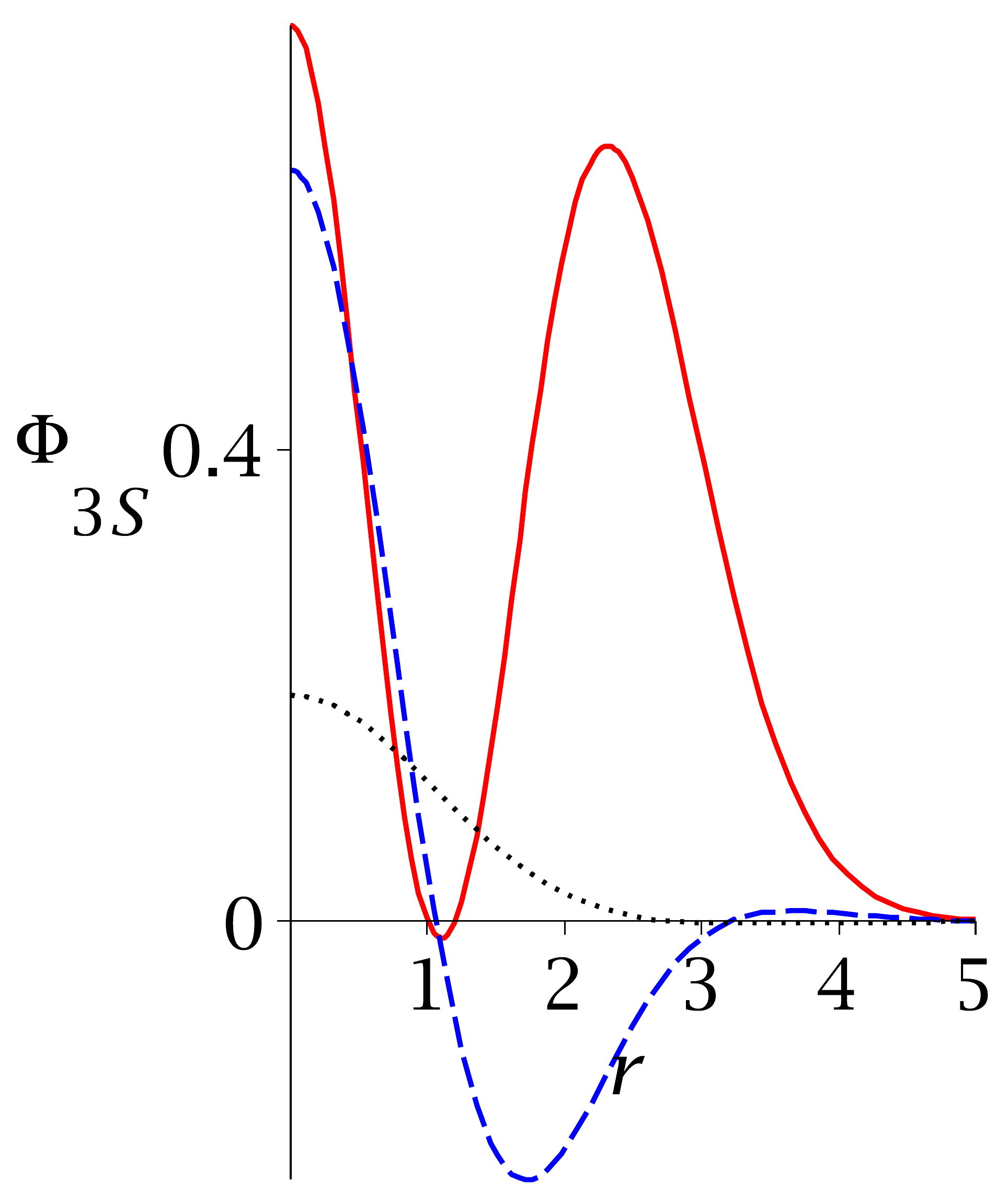}
 \caption{The scalar part of the light-cone wavefunction for
   $\Upsilon(1S)$ (left), $\Upsilon(2S)$ (center) and $\Upsilon(3S)$
   (right) as a function of the transverse size, $r$
   ($\mbox{GeV}^{-1}$),  of the $b\bar{b}$ pair with the light-cone
   momentum fraction carried by the quark, $z=0.5$ (solid), $z=0.3$ (dashed) and $z=0.2$ (dotted).}
 \label{fig:scalarwf}
\end{figure}

The vector meson wavefunction is less well known. Various models are
discussed in \cite{Forshaw:2004} and here we shall use the boosted
gaussian wavefunction, which works well for the light mesons, $\rho$ and
$\phi$ \cite{Forshaw:2004}, and also for the heavier $J/\Psi$ meson
\cite{Forshaw:2006np}. In this model, the meson light-cone
wavefunction is given by
\begin{align}
\psi_{\lambda,h\bar{h}}^{\Upsilon}(z,r)= &  \sqrt{\frac{N_{\text{C}}}{4\pi }%
}\frac{\sqrt{2}}{z(1-z)} \{\delta_{h,\bar{h}}\delta_{\lambda,2h}m_{b}%
\nonumber\\
&  +i(2h)\delta_{h,-\bar{h}}\text{e}^{i\lambda\phi}\left[
(1-z)\delta_{\lambda,-2h}+z\delta_{\lambda,2h}\right]\partial_{r} \} \phi_{nS}(z,r)~,
\end{align}
where $\phi_{nS}(z,r)$ is obtained by boosting a Schr\"odinger
gaussian wavefunction using the Brodsky-Huang-Lepage prescription
\cite{bhl:80}:
\begin{equation}
\phi_{\mathrm{sch}}(\mathbf{p}^2) \rightarrow
\phi_{nS}\left(\frac{\mathbf{k}^2 + m_{b}^2}{4 z(1-z)} -m_{b}^2
\right)
\end{equation}
and $\mathbf{p}$ is the relative three-momentum  between the quark and
antiquark in the meson's rest frame while $\mathbf{k}$ is their
relative transverse momentum in the boosted frame. The subscript $nS$ reminds us we are to
treat the different $\Upsilon$ states seperately. After a two-dimensional Fourier
transformation of the resulting light-cone wavefunction to
transverse $\mathbf{r}$-space, we obtain
\begin{equation}
 \phi_{nS}(r,z)=\left[\sum_{k=0}^{n-1} \alpha_{nS,k} \; R_{nS}^2 \; \hat{D}^{2k}(r,z) \right] G_{nS}(r,z)
\end{equation}
with $\alpha_{nS,0} = 1$.  The operator
\begin{equation}
 \hat{D}(r,z) = \frac{m_f^2 -\nabla_{r}^2}{4z(1-z)}-m_f^2
\label{operator-Laplacian}
\end{equation}
with $\nabla_{r}^2= \frac{1}{r} \partial_r+ \partial_r^2$, acts on the
Gaussian function
\begin{equation}
 G_{nS}(r,z)=\mathcal{N}_{nS}\; z(1-z) \exp \left(-\frac{m_b^{2}R_{nS}^{2}}{8z(1-z)}\right) \exp\left(-\frac{2z(1-z)r^{2}}{R_{nS}^{2}}\right)\exp\left(\frac{m_b^{2}R_{nS}^{2}}{2}\right).
\label{boosted-gaussian}
\end{equation}
Explicitly we obtain\footnote{We note that if the double partial derivative $\partial_r^2$ is used in 
Eq.~(\ref{operator-Laplacian}) instead of the Laplacian operator,
  $\nabla_{r}^2$, the $2S$ wavefunction reduces to that presented in \cite{nnpz:97}.} 
\begin{equation}
 \phi_{1S}(r,z)=G_{1S}(r,z) \;,
\end{equation}

\begin{equation}
 \phi_{2S}(r,z)=G_{2S}(r,z)[1+\alpha_{2S,1} \; g_{2S}(r,z)]
\end{equation}
and
\begin{equation}
 \phi_{3S}(r,z)=G_{3S}(r,z)\left \{ 1+\alpha_{3S,1}\; g_{3S}(r,z)+\alpha_{3S,2}\left[ g_{3S}^2(r,z)+4\left(1-\frac{4z(1-z)r^2}{R_{3S}^2}\right)\right] \right \}~,
\end{equation}
where
\begin{equation}
 g_{nS}(r,z)=2-m_{f}^2 R_{nS}^2 + \frac{m_{f}^2 R_{nS}^2}{4z(1-z)} -\frac{4z(1-z)r^2}{R_{nS}^2} \;.
\end{equation}
The scalar wavefunctions $\phi_{1S}$, $\phi_{2S}$ and $\phi_{3S}$ are
shown in Figure~\ref{fig:scalarwf} for $z=0.2,0.3$ and $0.5$. 

\begin{figure}
 \centering
\includegraphics[width=4.7cm]{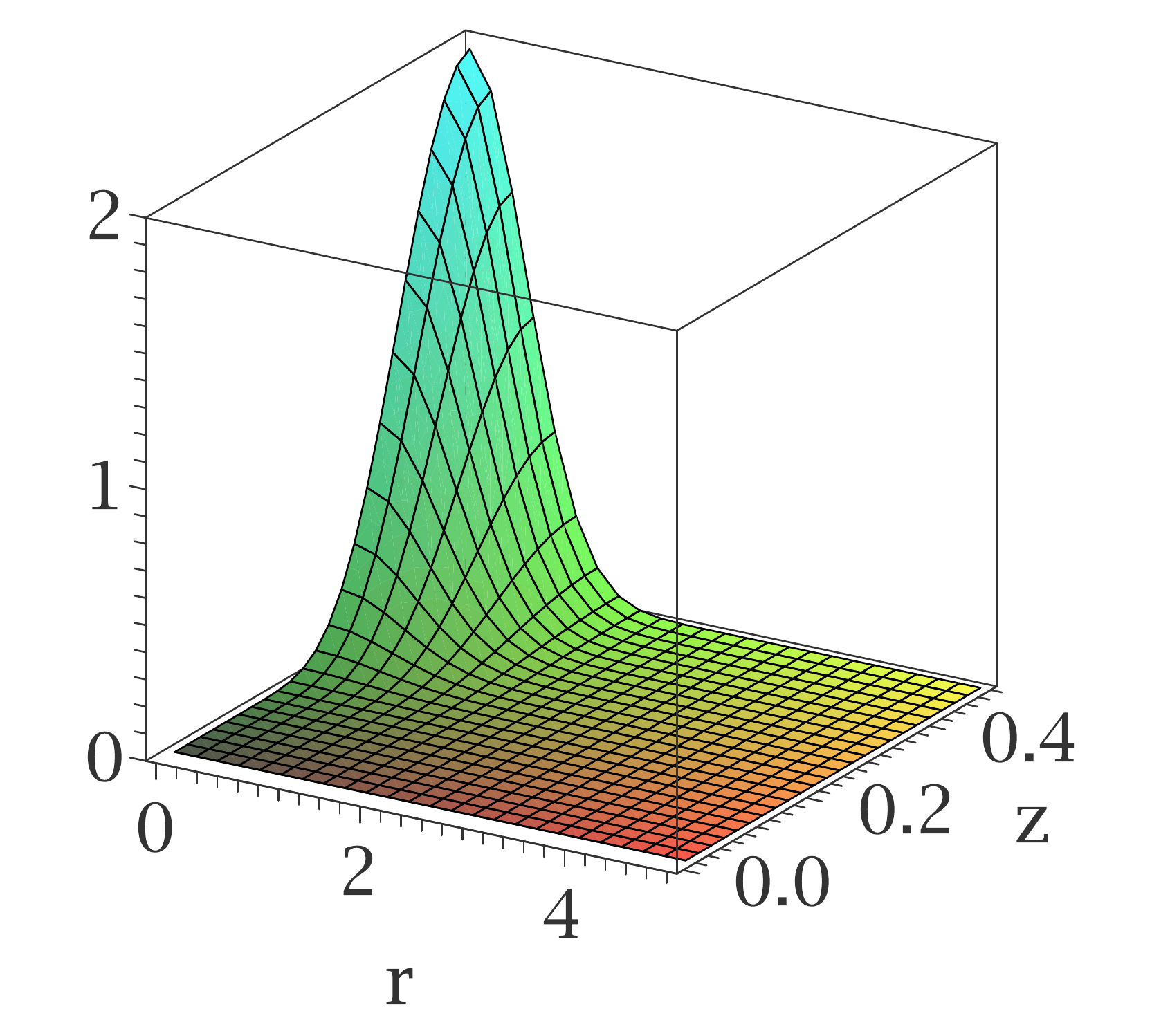}
\includegraphics[width=4.7cm]{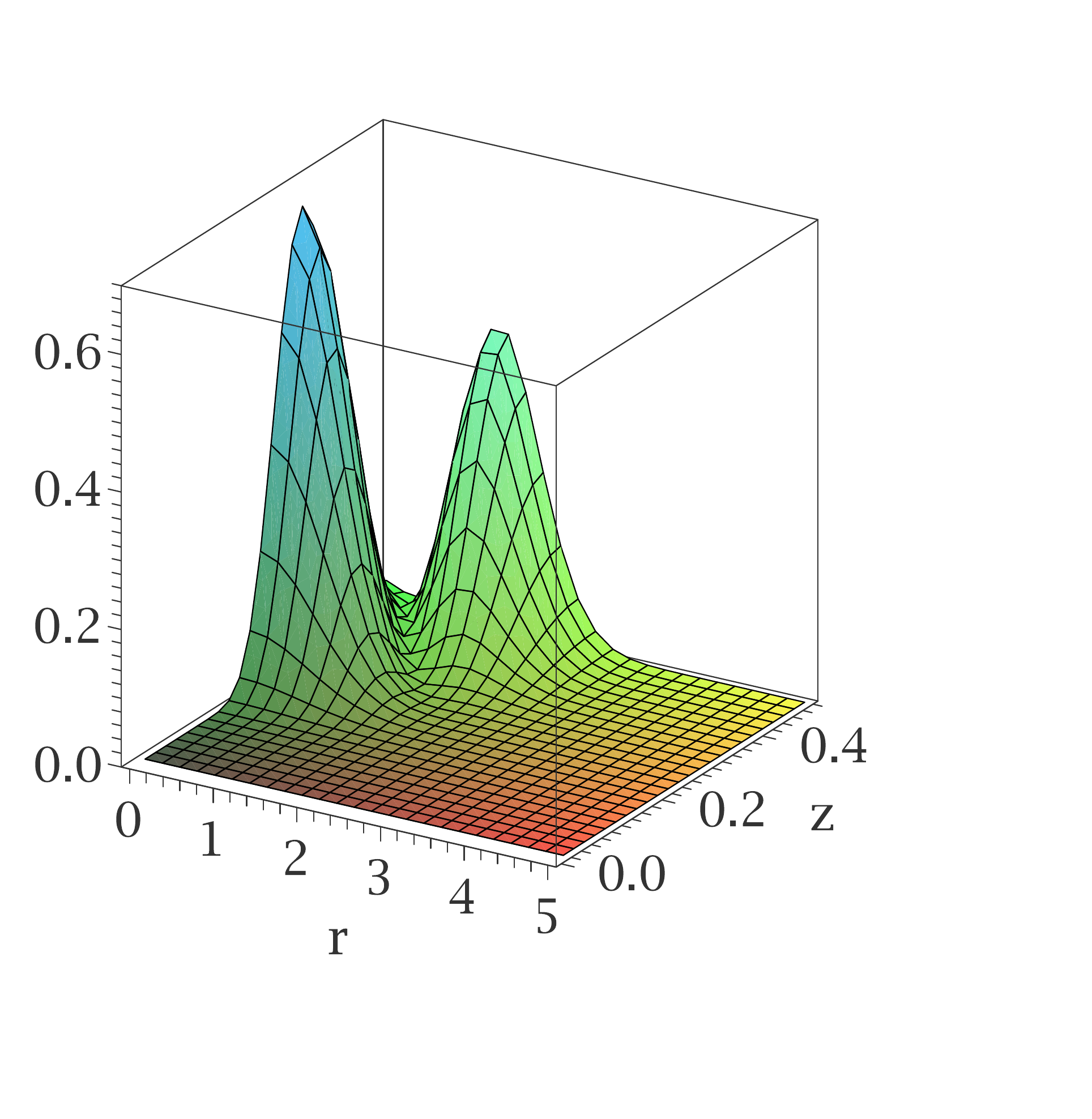}
\includegraphics[width=4.7cm]{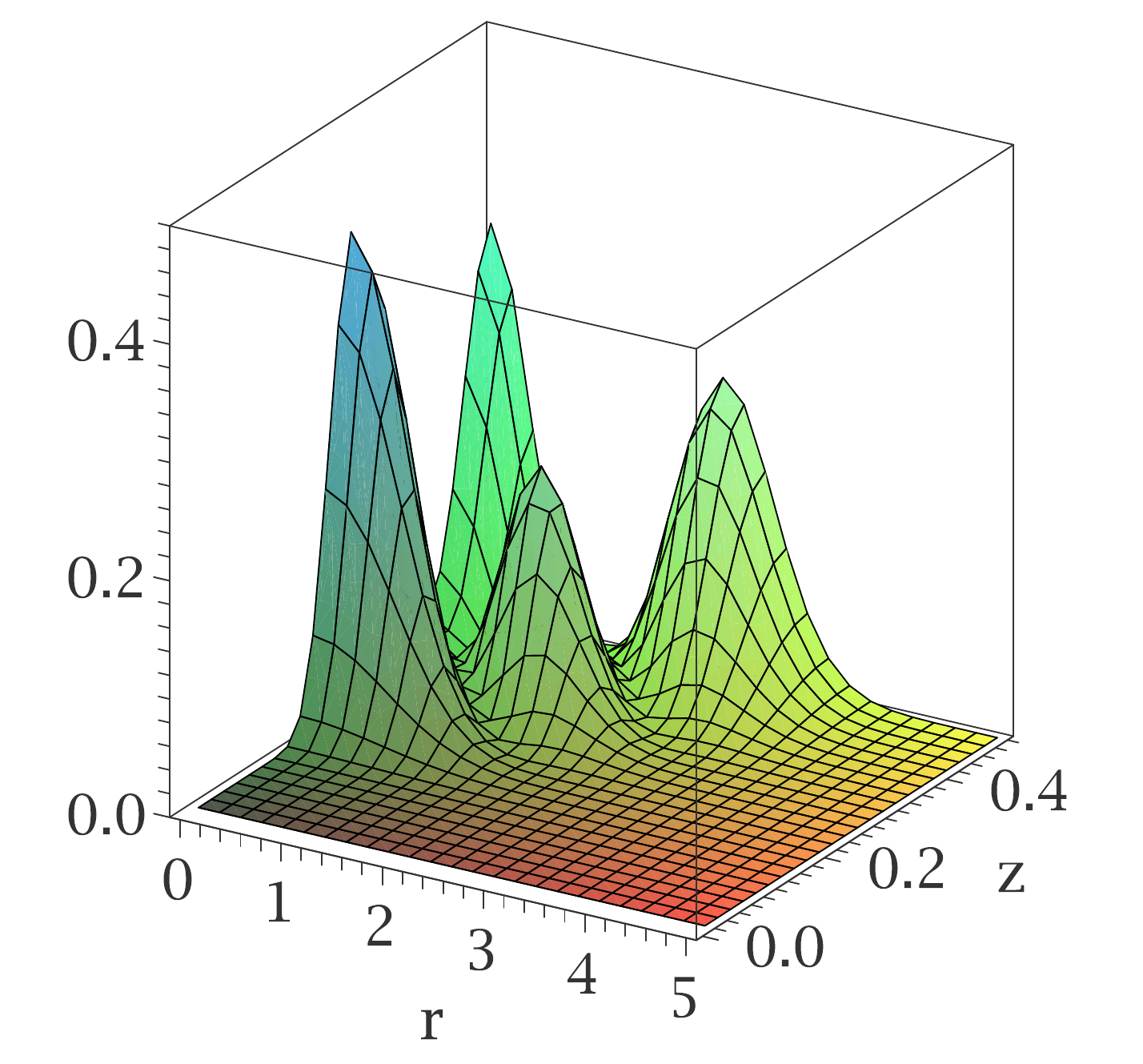}
 \caption{The light-cone wavefunction squared, $|\Psi^{\Upsilon}|^2$,
   for the transversely polarised $\Upsilon(1S)$ (left),
   $\Upsilon(2S)$ (center) and $\Upsilon(3S)$ (right) as a function of
   the transverse size, $r$ ($\mbox{GeV}^{-1}$),  of the $b\bar{b}$
   pair and the light-cone momentum fraction, $z$,  carried by the
   quark.}
\label{fig:upsilonwf}
\end{figure}

The parameters $\alpha_{nS,k}$ (for $k > 0$) have been fixed by requiring that the
wavefunctions of different states be orthogonal to each other. The
values of $R_{nS}$ are fixed so as to reproduce the experimentally measured electronic decay widths
\cite{Forshaw:2004}. The values of the parameters thus obtained,
together with the predicted decay widths, are given in Table
\ref{tab:gaussianwfn-params} and the resulting
wavefunctions are shown in Figure~\ref{fig:upsilonwf}.

\begin{table}[h]
\begin{center}
\textbf{Boosted Gaussian parameters}
\[%
\begin{array}
[c]{|c|c|c|c|c|c|c|}\hline
n & R_{nS}^{2} & \alpha_{nS,1} & \alpha_{nS,2} &\mathcal{N}_{nS} & \Gamma_{e^+ e^-} & \Gamma_{e^+ e^-}^{\mathrm{exp}}%
\\\hline
1 & 0.567 & . & .&0.481 &1.340& 1.340 \pm 0.018\\ \hline
2 & 0.831 & -0.555&. &0.624 &0.611& 0.612 \pm 0.011 \\ \hline
3 & 1.028  & -1.219& 0.217 &0.668 & 0.443 & 0.443 \pm 0.008 \\ \hline
\end{array}
\]
\end{center}
\caption{Parameters and predicted electronic decay widths of the Boosted Gaussian light-cone
wavefunctions in appropriate GeV based units. The decay widths are
given in keV. Experimental values are taken from \cite{pdg08} and we
take $m_b = 4.2~$GeV.}
\label{tab:gaussianwfn-params}
\end{table}

It remains to specify the parametric form of the dipole
cross-section. Here we use the saturation model presented in \cite{FS:2004}
(which we subsequently refer to as the ``FSSat'' model). It has the
following form:
 \begin{align}
\sigma(x,r) &  =A_{H}r^{2}x^{-\lambda_{H}}~~{\mathrm{for}}
~~r<r_{0}~~{\mathrm{and}}\nonumber\\
&  =A_{S}x^{-\lambda_{S}}~~{\mathrm{for}}~~r>r_{1}.\label{eq:FS04}
\end{align}
In the intermediate
region $r_{0}\leq r\leq r_{1}$, the dipole cross-section is determined by
interpolating linearly between the two forms of Eq.~(\ref{eq:FS04}).
$r_{0}$  is fixed to be the value at which the hard component is some
specified fraction of the soft component, i.e.
\begin{equation}
\sigma(x,r_{0})/\sigma(x,r_{1})=f~,
\end{equation}
where $f$ is treated as a parameter that is determined, along with the
other parameters $A_{S},A_{H},\lambda_{S},\lambda_{H}$ and $r_{1}$, by a fit
to the total structure function $F_{2}$ data from HERA \cite{FS:2004}.

We now have almost all of the ingredients we need in order to compute the
amplitude (\ref{factorized-amplitude}) for $\Upsilon$ production. The
forward differential cross-section is given by
\begin{equation}
\left.  \frac{\text{d}\sigma_{\gamma p}}{\text{d}t}\right\vert _{t=0}=\frac
{1}{16\pi} \vert \mathcal{A}_{\gamma p} \vert^2
\label{gammap-difxsection}
\end{equation}
and the total cross-section is
obtained by assuming an exponential fall-off with increasing $|t|$, i.e
\begin{equation}
 \sigma_{\gamma p}  = \frac{1}{B_{\Upsilon}}
 \left.\frac{\text{d}\sigma_{\gamma p}}{\text{d}t}\right|_{t=0}~,
\label{gammap-xsec}
\end{equation}
where the diffractive slope, $B_{\Upsilon}$, is taken to be \cite{Forshaw:2004}
\begin{equation}
B_{\Upsilon}=N\left(
  \frac{14.0}{(M_{\Upsilon}/\text{GeV})^{0.4}}+1\right) = 3.68~\mathrm{GeV}^{-2}
\label{Bslope}
\end{equation}
with $N=0.55$ GeV$^{-2}$. 
Before presenting our predictions for hadron-hadron collisions, we
shall first compute the photon-proton cross-section $\sigma_{\gamma
p}$ and compare it to existing HERA data.

\section{The $\gamma p$ cross-section}
\begin{figure}
 \centering
\includegraphics[width=10cm]{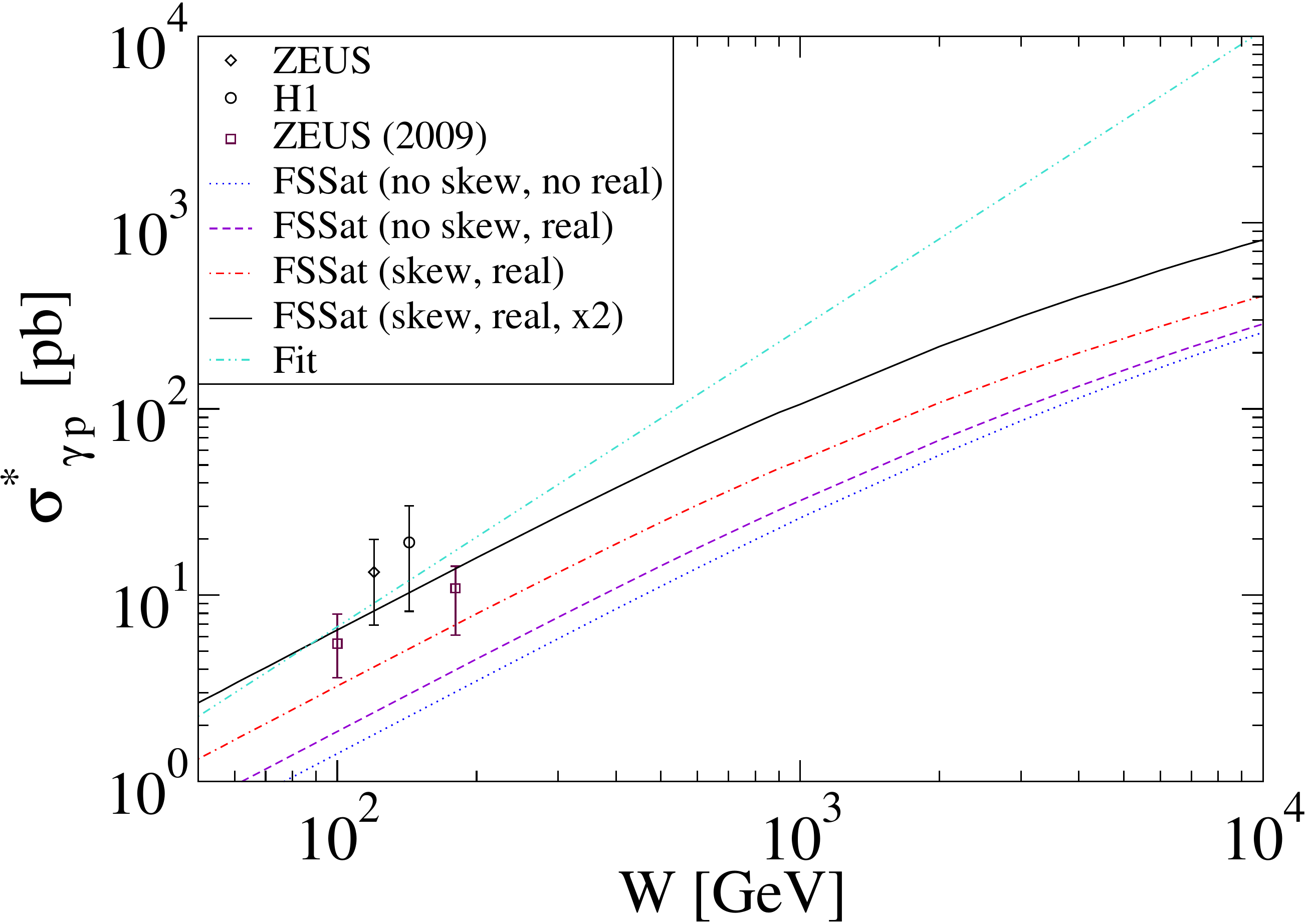}
 \caption{Predictions for the $\gamma p$ cross-section. Dotted
   curve -- no skewedness correction and without the real part;
   dashed curve -- including the real
   part; dot-dashed curve -- including the real part and a skewedness
   correction. The solid curve is the dot-dashed curve normalised to the HERA data
   and the double-dot-dashed curve is the parameterization to the HERA data
   described in the text \cite{MW:2008}.}
\label{fig:photon-proton-xsec}
\end{figure}

To compare to the data, we compute
\begin{equation}
 \sigma^{*}_{\gamma p}=\sum_{n=1}^3  \sigma_{\gamma p \rightarrow
   \Upsilon(nS) p}  \times B_{n}~,
\label{unresolved-gammapxsec}
\end{equation}
where $B_{n}$ is the branching ratio of the $\Upsilon(nS)$ to
muons. In Figure~\ref{fig:photon-proton-xsec}, we compare our
predictions to the HERA data \cite{ZEUS,H1,ZEUS:09}.

As can be seen, the theory curve (dotted) lies significantly below the
HERA data. However, we have ignored two important
corrections. The first is the contribution from the real part of the
amplitude and including it increases the cross-section for each state by a
factor of $(1+\beta^2)$ where $\beta$ is the ratio of the real to
imaginary part of the $\gamma p$ amplitude, given by 

\begin{equation}
 \beta=\tan \left(\frac{\pi}{2} \lambda \right) 
\end{equation}
with
\begin{equation}
 \lambda=\frac{\partial \ln |\Im \text{m}\; \mathcal{A}|}{\partial \ln \left(1/x\right)}
\label{logderivative}
\end{equation}
where $\Im \text{m}\; \mathcal{A}$ is given by Eqn.~(\ref{factorized-amplitude}).

\begin{figure}
 \centering
\includegraphics[width=7.5cm]{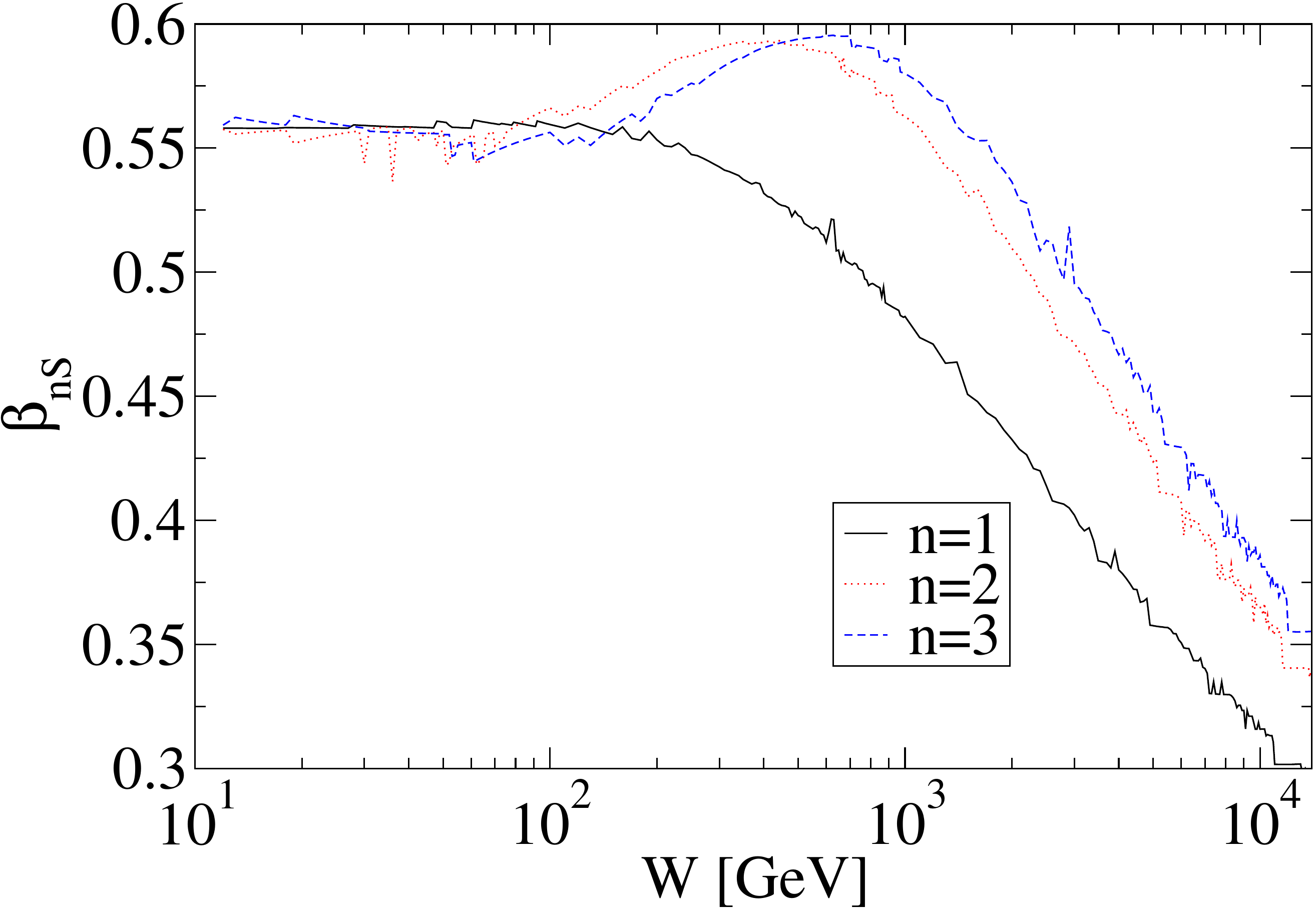}
 \caption{The ratio of the real to imaginary part of the amplitude for
   the different states.}
 \label{fig:beta}
\end{figure}

Figure \ref{fig:beta} shows the ratio of the real to imaginary parts
of the amplitude for each state. We note that the calculation of this
ratio is model-dependent especially for the higher states. The dashed
curve in Figure~\ref{fig:photon-proton-xsec} shows the effect of
including the correction due to the real part. 

The second correction arises because the
amplitude is not diagonal -- the mass of the $\Upsilon$ is large
and timelike compared to the small, spacelike, photon
virtuality. In contrast, our dipole scattering cross-section was
extracted from data at zero momentum transfer. We estimate the corrections from this source by
multiplying the amplitude by a factor of \cite{Shuvaev:1999}
\begin{equation}
R_{g} (\lambda) = \frac{2^{2 \lambda +3}
}{\sqrt{\pi}}\frac{\Gamma(\lambda + 5/2) }{\Gamma(\lambda + 4)}~.
\end{equation}
Our prediction, with this correction included, is shown in
Figure~\ref{fig:photon-proton-xsec} as the dot-dashed curve.

A few comments are in order regarding our estimation of the above two
corrections. Both of them depend on $\lambda$, which we estimate using
Eqn.~(\ref{logderivative}). We have checked that our
results do not vary very much upon choosing a fixed value of $\lambda
= 0.3$. Since the $\Upsilon$ wavefunction extends to larger $r$ for
the excited states, relative to that of the $1s$ state
(see Figure \ref{fig:upsilonwf}), one might naively expect that the
energy dependence is softer for the former, resulting in a
lower effective value of $\lambda$. This is actually not the case as
one can infer from Figure~\ref{fig:beta}. This is because the excited
state wavefunctions have nodes, which leads to a partial cancellation
between the soft and hard parts of
the amplitude which in turn results in an effective harder energy dependence
(higher $\lambda$) in the amplitudes. The degree
of cancellation is rather sensitive to the model used for the dipole
cross-section.  This has been referred to previously as the node effect \cite{nnpz:97}.  

Even with these corrections, the theory curve still lies below
the data. This is not a particularly surprising result and it is
typical of other dipole model calculations
\cite{MW:2008, ZEUS:DIS08}. Quite possibly, one could do better by
improving the vector meson wavefunctions and in that case the
uncertainty would mainly influence the normalisation of the
cross-section. Moreover, the value chosen for the $b$ quark mass also
affects the overall normalisation. For these reasons, we follow the
authors of \cite{MW:2008} and rescale our results (by a factor $2.0$
with $m_b = 4.2$~GeV) in order to achieve optimum agreement with the
data. The result is the solid curve in
Figure~\ref{fig:photon-proton-xsec}. By doing this, we account also
for the uncertainties in the real part and skewedness corrections in a
purely phenomenological way. 

In \cite{MW:2008}, the parametrisation 
\begin{equation}
\sigma = 0.12~\text{pb}~\times \left( \frac{W}{\text{GeV}}
\right)^{1.6} 
\label{fit}
\end{equation}
was found to fit the HERA data for $\Upsilon(1S)$. Both H1 and ZEUS
measure $\sigma_{\gamma p}^*$
and then extract the $1S$ cross-section by assuming that the ratios
of cross-section times branching ratio are the same as those measured
by the CDF collaboration \cite{CDF:95}, i.e. 
$(\sigma_{2S}\cdot B_{2S})/(\sigma_{1S}\cdot B_{1S})=0.281 \pm 0.0484$
and $(\sigma_{3S}\cdot B_{3S})/(\sigma_{1S}\cdot B_{1S})=0.155 \pm 0.0319$. 
We use Eq.~(\ref{fit}), together with the CDF ratios,
to produce the double-dot-dashed curve in
Figure~\ref{fig:photon-proton-xsec}.

\begin{figure}
 \centering
\includegraphics[width=7.5cm]{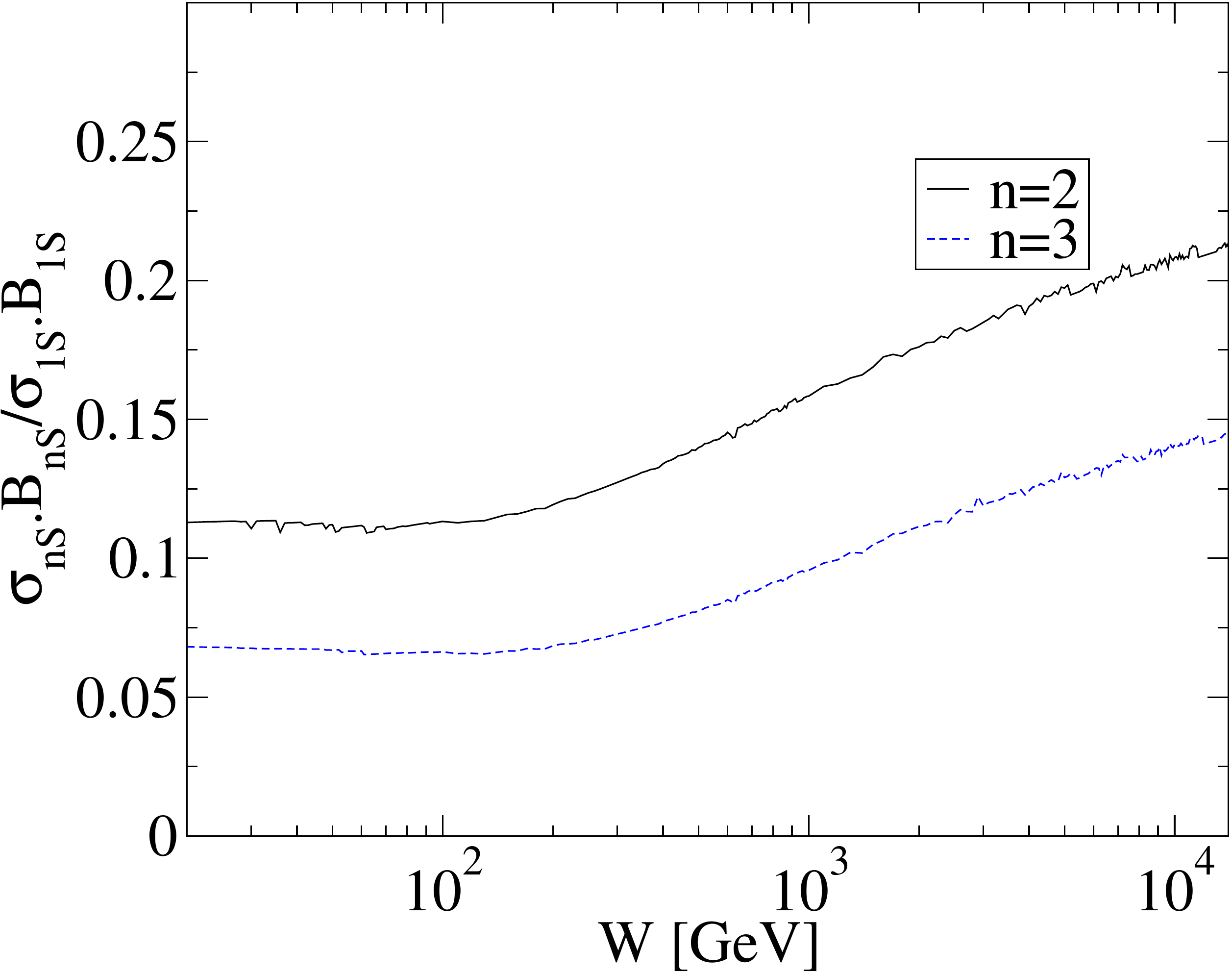}
 \caption{The ratio of cross-sections for the different states. }
 \label{fig:ratio}
\end{figure}

We can also compute the cross-section ratios between the different $\Upsilon$
states in order to compare to the CDF data. These are shown in
Figure~\ref{fig:ratio}. Our results are below the values obtained by
CDF quoted above.  For the $2S:1S$ ratio, in the range $50 < W < 200$
GeV,  our prediction is also below the value calculated in
\cite{RSS:2008} using perturbative QCD and a Gaussian
wavefunction. However, theoretical predictions for these ratios are rather uncertain since the $2S:1S$ ratio is
very sensitive to the details of the meson wavefunction and also to the value of $\lambda$.

Note that the FSSat model predicts an energy
dependence that is less steep than the $W^{1.6}$ dependence of the
fit. We might take the difference to be indicative of
the challenge facing future experiments, i.e. they should be able to
distinguish between the two.

\section{Results}
We are now ready to present our predictions for  the hadroproduction
cross-section. We assume that the angular distribution of the decay muons in the $\Upsilon$
rest frame is $\propto (1 + \cos^2 \theta^*)$, which corresponds to a
distribution of 
\begin{equation}
 \frac{\mathrm{d}^2N}{\mathrm{d}(\cos\theta)\mathrm{d}\phi} = \frac{3}{16\pi}
 \frac{1-\beta^2}{(1-\beta \cos\theta)^2}\left[ 1+\left(\frac{\cos
       \theta - \beta}{(1-\beta\cos \theta)}\right)^2 \right]
\end{equation}
in the lab: $\beta = \tanh Y$ is the speed of the $\Upsilon$ and $\theta$ is the
polar angle relative to the direction in which the  $\Upsilon$ travels
when $\beta > 0$.\footnote{Recall that we neglect the transverse
  momentum of the meson, which means it travels along the beam axis.} Thus
\begin{equation}
 \frac{\text{d}^2\sigma(pp\rightarrow p \{\mu^+ \mu^- \}
     p)}{\text{d}(\cos \theta) \text{d}Y } =  \frac{\mathrm{d}N}{\mathrm{d}(\cos\theta)} \xi \sigma_{\gamma
   p}(\xi s) \; f_{\gamma/ p}(\xi) \; + (Y \to -Y)~.
\end{equation}
We integrate over $\theta$ such that both the $\mu^-$ and $\mu^+$ are
emitted at an angle greater than $\theta_{\mathrm{min}}$ relative to
the beam. To do that we need $\alpha$, the emission
angle of the $\mu^+$:
\begin{equation}
 \cos \alpha = \frac{(1+\beta^2) \cos \theta - 
2\beta}{1+\beta^2-2\beta\cos \theta}~.
\label{alpha-theta}
\end{equation}
The transverse momentum of the muon (or anti-muon) is given by
\begin{equation}
 P_{t}=\frac{M_{\Upsilon}^2 \sin \theta}{2E_\Upsilon(1-\beta \cos \theta)} 
\end{equation}
and we also cut on a minimum value of $P_{t}^{\mathrm{min}}$ for both leptons.

For the Tevatron, we approximate the angular acceptance of the CDF
muon detector using $\theta_{\mathrm{min}}= 33.5^\circ$, whilst for the D\O\ detector we take
$\theta_{\mathrm{min}}=15.4^\circ$. At the LHC we assume,
$\theta_{\mathrm{min}}=7.7^\circ$ (see also \cite{ovyn}). In all cases we show results for
$P_{t}^{\mathrm{min}} = 3$~GeV and $P_{t}^{\mathrm{min}}=4$~GeV.
At the LHC, we may also have the possibility to detect one of the
protons in the region 420~m from the interaction point (there will not be any acceptance to detect both). According to the studies presented in \cite{Higgs1}, forward detectors in the 420~m region have acceptance for protons with fractional energy loss $0.002 < \xi < 0.018$ on the right-hand side and $0.0015
< \xi < 0.014$ on the left-hand side.

\begin{figure}
\centering
\includegraphics[width=7cm]{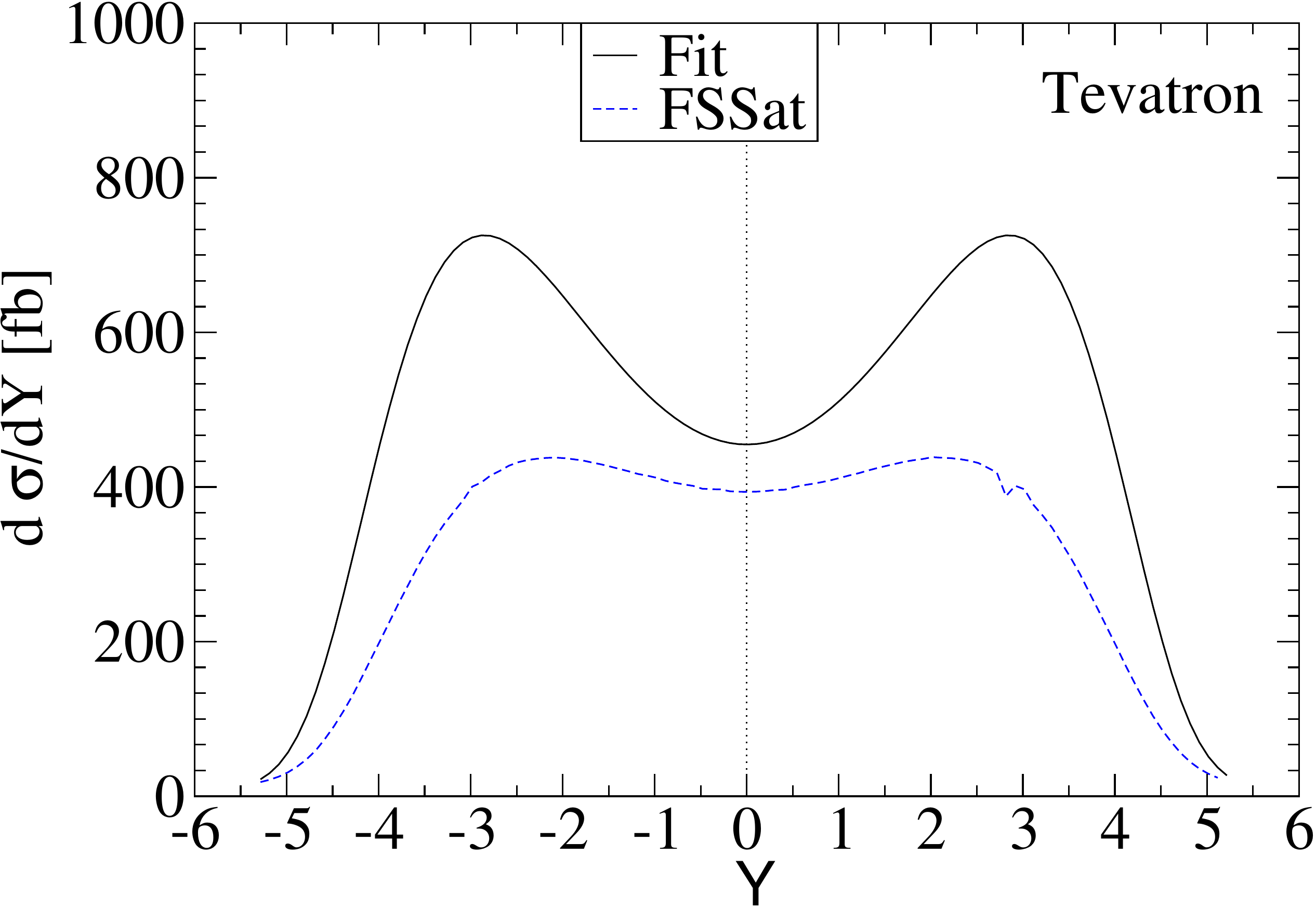}
\hspace{1cm}\includegraphics[width=7cm]{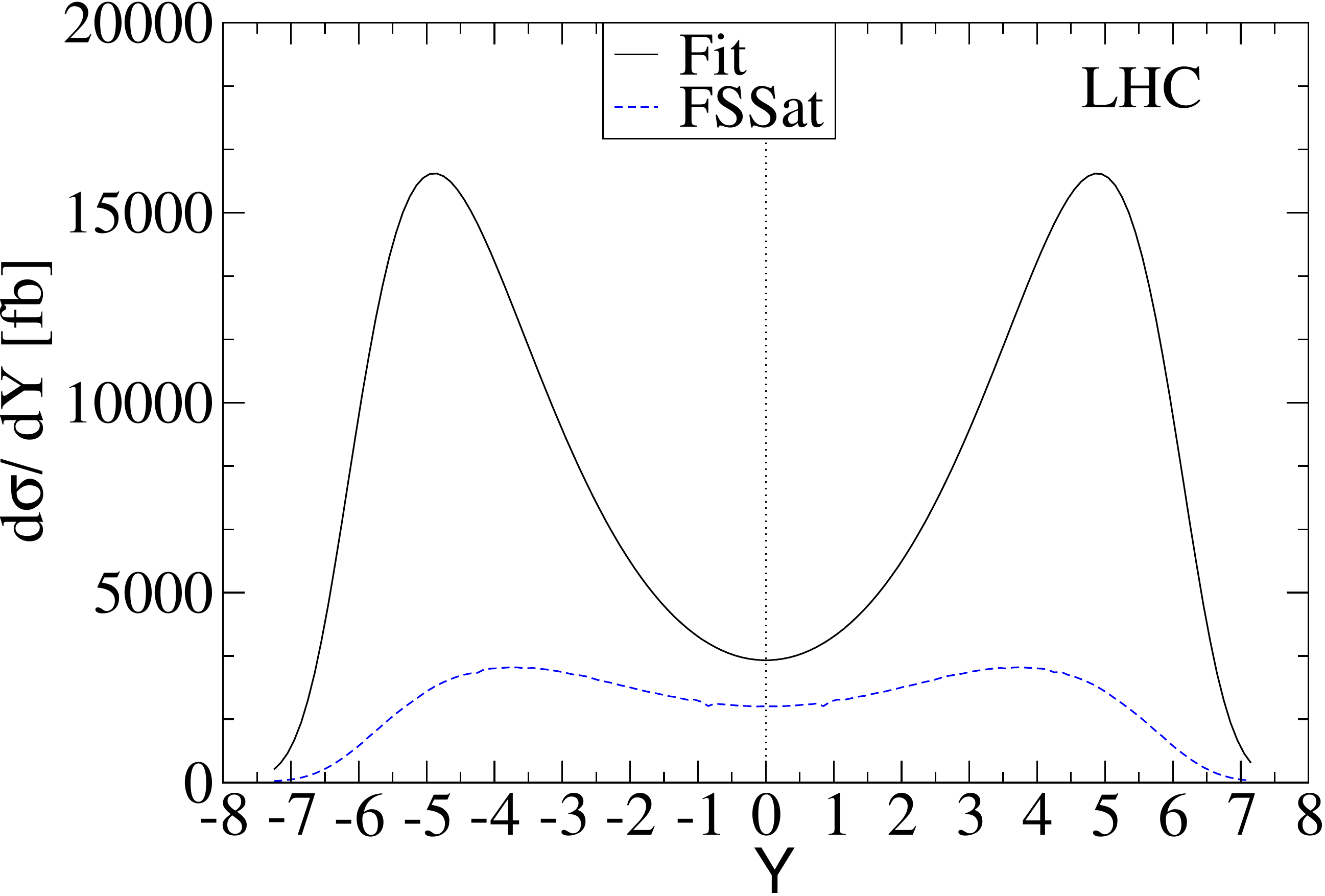}
\caption{The $\Upsilon$ rapidity
  distribution without any cuts on the final state particles at the
  Tevatron (left) and LHC (right) using the FSSat dipole model and the
  parameterization the photoproduction data described in the
  text.}
 \label{fig:rapdist-nocuts}
\end{figure}

\begin{figure}
\centering \vspace*{1cm}
\includegraphics[width=7cm]{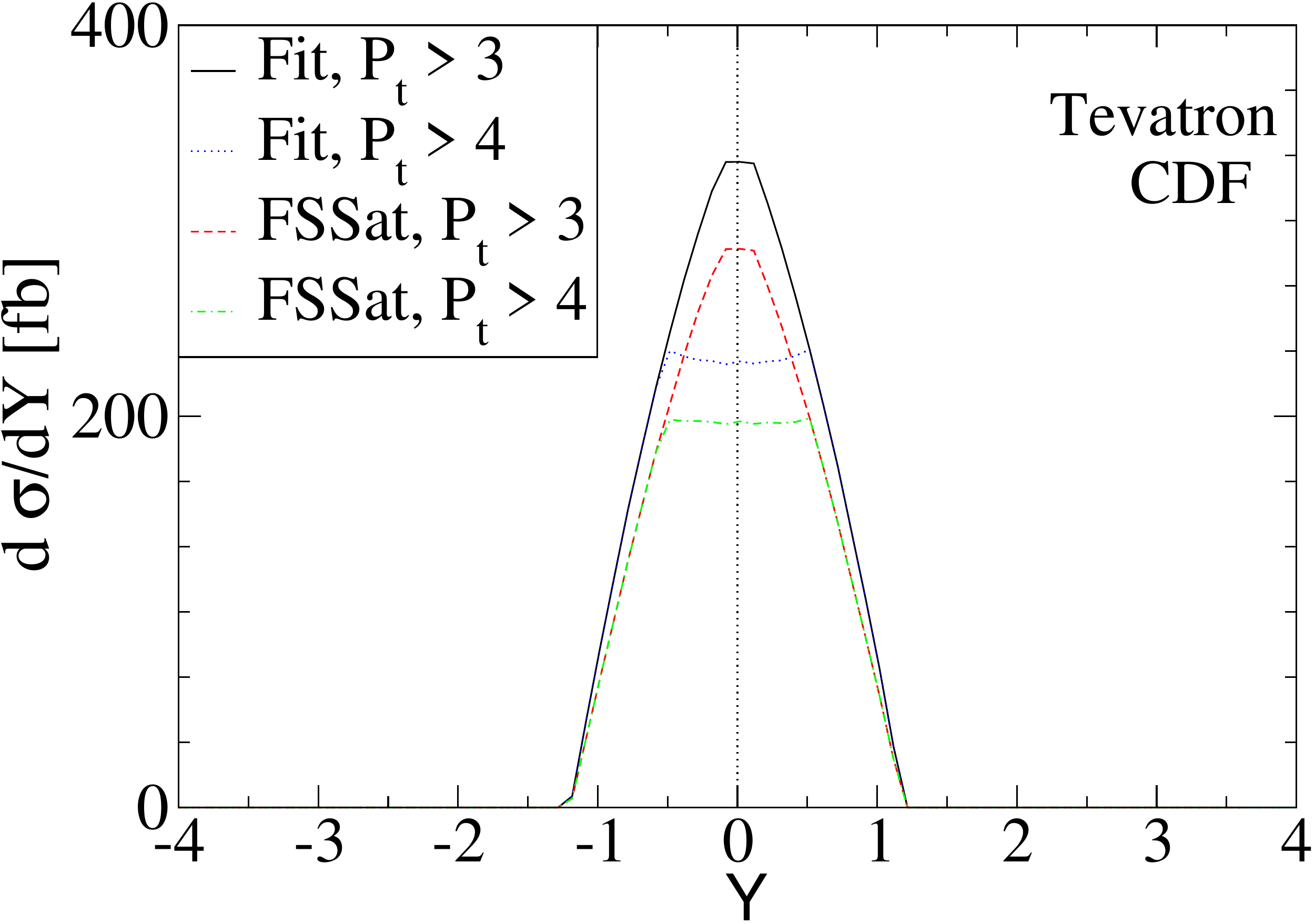}
\hspace{1cm}\includegraphics[width=7cm]{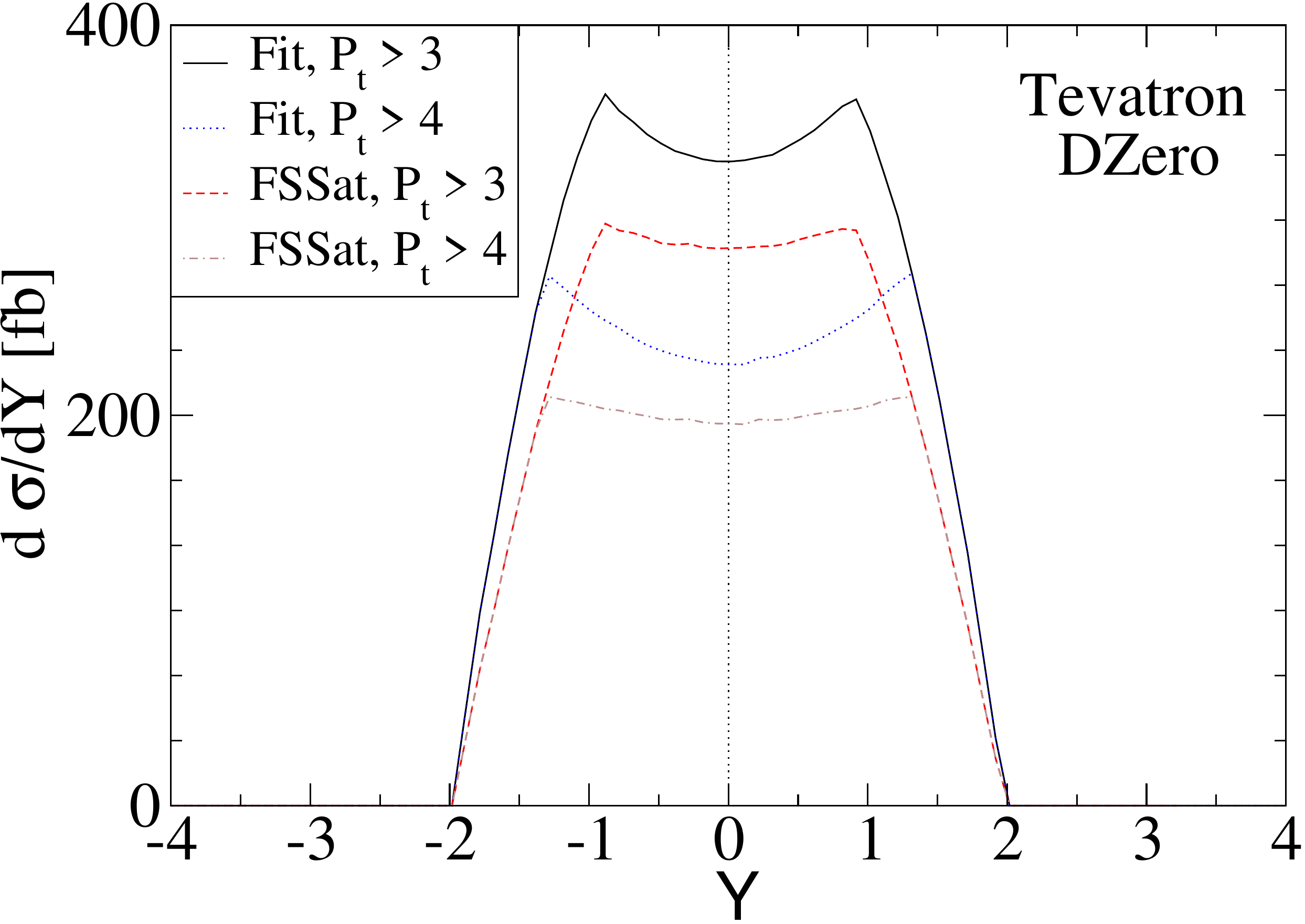}\\ \vspace*{1.0cm}
\includegraphics[width=8cm]{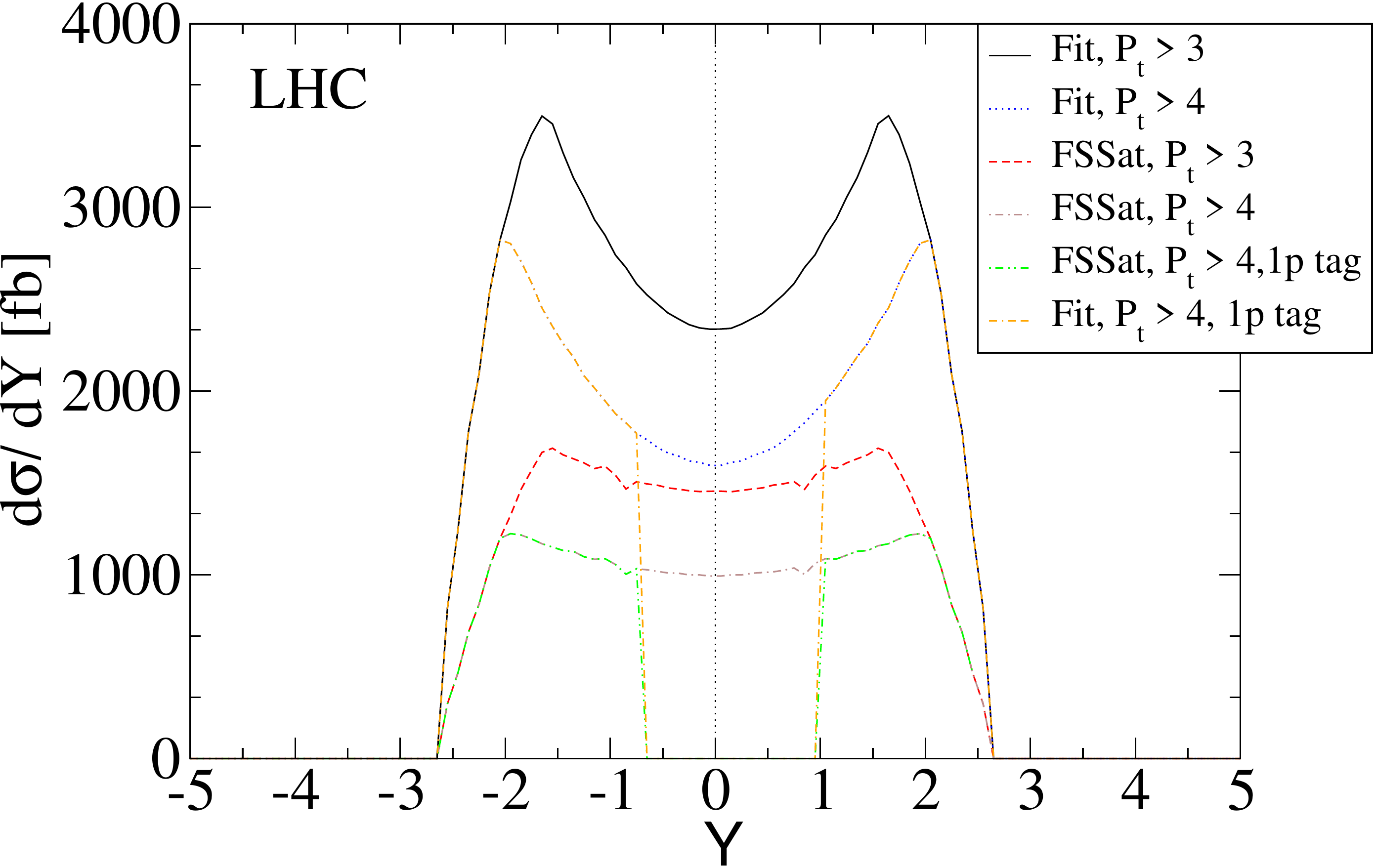}
\caption{The $\Upsilon$ rapidity
  distribution after cuts at the Tevatron (top) and LHC (bottom) using the FSSat dipole model and
  the the parameterization the photoproduction data described in the
  text.}
 \label{fig:rapdist-cuts}
\end{figure}

Figure~\ref{fig:rapdist-nocuts} shows the $\Upsilon$ rapidity
distributions at the LHC and the Tevatron respectively before any
cuts have been applied. There is a
striking difference (especially at the LHC) between the shapes of the
distributions obtained using the the FSSat dipole model and the
parameterization of the photoproduction cross-section that is based on the
HERA data. One could therefore hope  that measurements
of the rapidity distributions would be able to discriminate between
dipole models and thereby contrain the gluon
distribution. In Figure~\ref{fig:rapdist-cuts} we indicate the effect
of cutting on the muon $p_t$ and rapidity and, in the case of the LHC,
the effect of observing one of the protons in the proposed 420~m detectors.
Our results show that although the strong
sensitivity to the energy dependence on the dipole cross-section is
diminished, it is still large enough that one could hope
to constrain the theory.

\begin{table}[h]
\begin{center}
\textbf{Cross-sections at the Tevatron}
\[
\begin{array}
[c]{|c|c|c|}\hline
\mbox{} &\sigma_{\mbox{\tiny{FSSat}}}(\mbox{fb}) & \sigma_{\mbox{\tiny{Fit}}}(\mbox{fb}) \\ \hline
\mbox{CDF}& 350 \;  &  406\\ \hline
\mbox{D\O}& 685 \;  & 838 \\ \hline
\end{array}
\]
\end{center}
\caption{The predicted cross-sections for the Tevatron using the FSSat
  dipole model (first column) and the power-law fit to the
  photoproduction data (second column). Cuts on the muon rapidities
  and transverse momentum have been applied ($P_{t}> 4$ GeV).} 
\label{tab:ppxsections-tevatron}%
\end{table} 
\begin{table}[h]
\begin{center}
\textbf{Cross-sections  at the LHC}
\[%
\begin{array}
[c]{|c|c|c|}\hline
\mbox{No. of tagged protons} & \sigma_{\mbox{\tiny{FSSat}}}(\mbox{fb}) & \sigma_{\mbox{\tiny{Fit}}}(\mbox{fb})\\ \hline
0& 5133  & 10351 \\ \hline
1& 3035 & 6802 \\ \hline
\end{array}
\]
\end{center}
\caption{The predicted cross-sections for the LHC using the FSSat
  dipole model (first column) and the power-law fit to the
  photoproduction data (second column). Cuts on the muon rapidities
  and transverse momentum ($P_{t}> 4$ GeV) have been applied (first row). The result of
  additionally measuring one of the outgoing protons is shown in the
  second row.}
\label{tab:ppxsections-LHC}%
\end{table} 

Tagging one of the protons does severely limit the
acceptance of a measurement: Centrally produced mesons are cut out because the 420~m
detectors select events in which the measured proton loses much more
energy than the unmeasured proton. The boost is so large in fact that
the upsilon always travels in the direction of the tagged proton.
However, detecting a proton does have the potential advantage that the
measurement need not only be performed using data collected at low
($\lesssim 10^{32}\lumi$) luminosities. With one tagged proton, one could
hope to control the pileup background after utilising the fact that
the two charged muons should point to a single vertex (and no other
tracks point to the same vertex) and that they
should also combine to produce an $\Upsilon$ rapidity that is in agreement
with the value inferred from the extracted (one measured directly and
one inferred) proton momentum fractions. Further detailed simulations would be required to establish whether this method will 
work at $2 \times10^{33}\lumi$ and above. In any case, it is worth
noting that $10^{32}\lumi$ corresponds to 1~fb$^{-1}$ per year. That
translates to a total of over 5000 signal events in an environment
where there will be, on average, less than one $pp$ interaction per bunch crossing. This should be sufficient
to produce a measurement that will constrain QCD models of saturation. 

Finally, it is worth pointing out that should it be possible to
make a cut on the measured proton's transverse momentum then it would
become possible to make a direct measurement of the
photoproduction cross section. This is so since the transverse momentum of
protons that radiate photons is typically much smaller than
the transverse momentum of protons that do not, i.e. Eq.~(\ref{photon-flux}) is much
softer than the $\exp(B_\Upsilon t)$ dependence implied by
Eq.~(\ref{Bslope}). Requiring that the transverse momentum of the tagged 
proton $p_T < 100$ MeV is at least
60\% efficient for selecting events in which the tagged proton radiates a
photon\footnote{The efficiency depends weakly upon the energy lost by the
  proton through Eq.~(\ref{Q2min}).}. The contamination from events in which the non-tagged 
  proton radiates a photon would be only 4\%. 
  If the cut is lifted to 300~MeV, the ratio decreases to
  $89\%:28\%$, which still constitutes a very significant
  enhancement. 

\begin{figure}
\centering
\includegraphics[width=9cm]{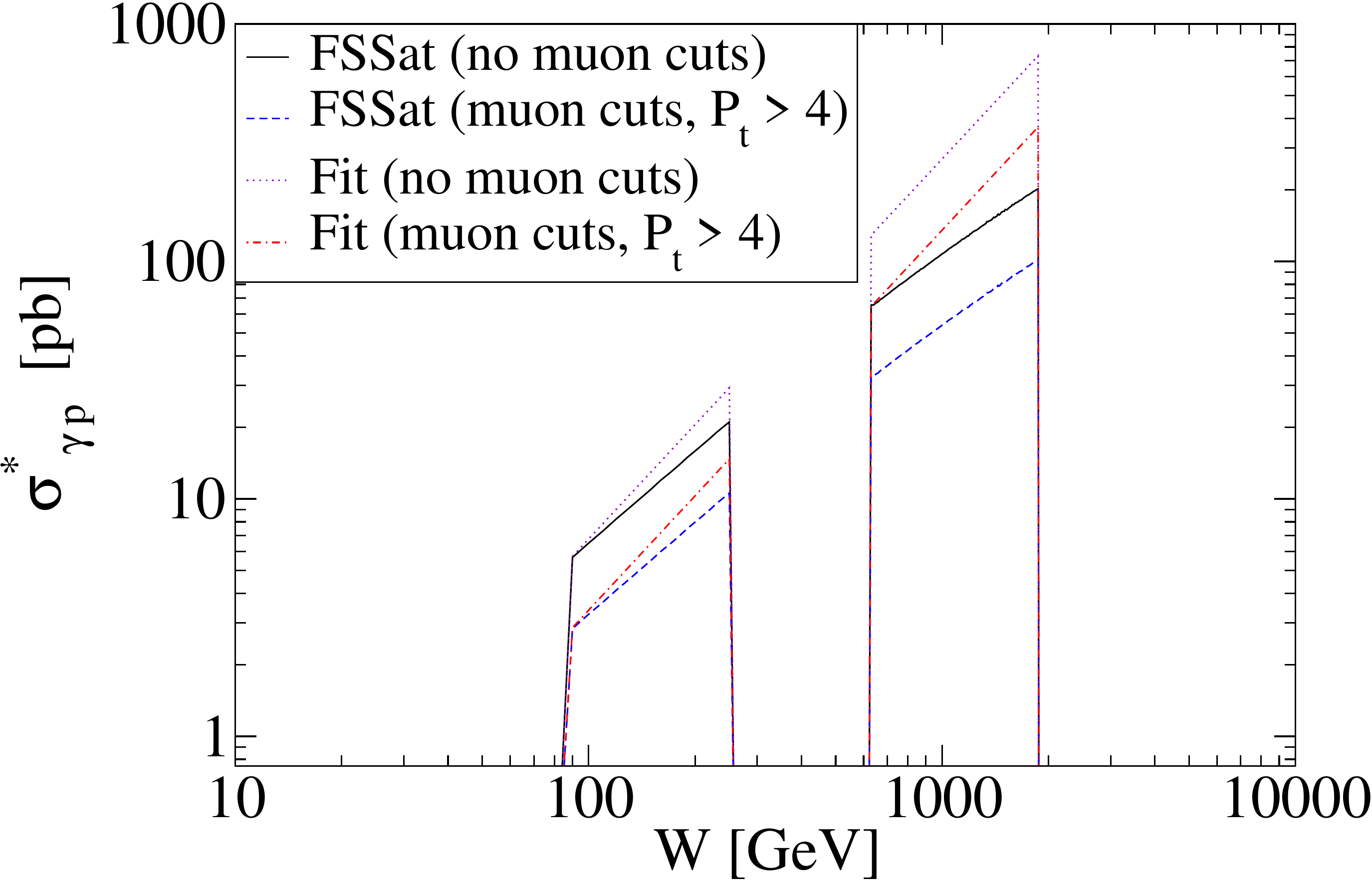}
\caption{The photoproduction cross-section extracted after cutting on
  the transverse momentum of the tagged proton.}
 \label{fig:gammap-cuts}
\end{figure}

  After making such a cut, since we now have an enriched sample of
  events in which the tagged proton radiated the photon, we can
  eliminate the photon flux and extract the photoproduction
  cross-section. Figure~\ref{fig:gammap-cuts} illustrates the
  possibilities. The region at larger $W$ (much larger than can be
  probed at HERA and the Tevatron) is obtained by insisting that the
  tagged proton $p_T$ be sufficiently small, in which case the tagged proton
  most likely radiated the photon. The lower $W$ region can be measured by making
  the complimentary cut, i.e. by insisting that the proton have $p_T$
  above some value. This is equivalent to assuming that the untagged proton emitted the photon. 
  In this way, the proton
  detectors facilitate a measurement of the $\gamma p \to \Upsilon p$
  cross-section at $W = 1$~TeV, which is well in the range where
  saturation effects are expected to reveal themselves.   

So far, we have made no mention of the issue of ``gap survival'' and
as far as the overall rate is concerned it is not expected to provide
much suppression, since the collision is typically rather
peripheral. However, the gap survival should depend rather strongly on
the proton $p_T$, with gaps being filled in more often at larger
values of $p_T$. Indeed measuring the $p_T$ distribution has been
advertised as a means to probe the gap-filling mechanism. This physics
would need to be under control before one could reliably extract the
photoproduction cross-section by exploiting a $p_T$ cut on the scattered proton  
\cite{Khoze:2002dc,Khoze:2008cx}.

\section*{Acknowledgements}

We wish to thank Thorsten Wengler, Andrew Pilkington and Graeme Watt
for helpful discussions. We also thank the Royal Society and the STFC
for financial support.

\end{document}